\documentclass[12pt]{article}
\usepackage[latin1]{inputenc}
\usepackage{graphicx}
\usepackage{color}
\usepackage{geometry}  % fuer Layout/Raender 
\usepackage{parskip}   % Abstand nach Absatz/keine Einrueckung
\usepackage{amsmath, amssymb} % fuer Formeln/Symbole
\usepackage{setspace} % fuer Zeilenabstand 
\usepackage{booktabs} % fuer Verweise 
\usepackage[round]{natbib} % fuer Literaturverzeichnis
\DeclareMathOperator{\E}{\mathbb{E}}
\usepackage{float}
\usepackage[misc]{ifsym}
\usepackage{tabularx}
\usepackage{subcaption}

\newcommand{\bs}[1]{\ensuremath{\boldsymbol{#1}}} % neuer Befehl fuer fette Zeichen
\DeclareMathOperator*{\argmax}{arg\,max}

\newcolumntype{Y}{>{\centering\arraybackslash}X}

\begin{document}

% Infos fuer den Dokumentenkopf
\title{Confidence intervals for tree-structured varying coefficients} 
\author{Nikolai Spuck\footnote{Institute of Medical Biometry, Informatics and Epidemiology, Medical Faculty, University of Bonn, Venusberg-Campus 1, 53127 Bonn, Germany.}\ , Matthias Schmid$^{*}$, Malte Monin\footnote{Department of Internal Medicine I, University Hospital Bonn, Venusberg-Campus 1, 53127, Bonn, Germany.}\ \footnote{German Centre for Infection Research (DZIF), partner-site Cologne-Bonn, Bonn, Germany.}\ ,\\ and Moritz Berger$^{*}$}
\date{\today}

% Kopf des Dokuments 
\maketitle
\vspace{-0.5cm}
% Abstract 
\begin{abstract}
\noindent The tree-structured varying coefficient model (TSVC) is a flexible
regression approach that allows the effects of covariates to vary with the values
of the effect modifiers. Relevant effect modifiers are identified inherently using
recursive partitioning techniques. To quantify uncertainty in TSVC models, we propose a
procedure to construct confidence intervals of the estimated partition-specific coefficients. This task constitutes a selective inference problem as the coefficients of a TSVC model result from data-driven model building. To account for this issue, we introduce a parametric bootstrap
approach, which is tailored to the complex structure of TSVC. Finite sample properties, particularly coverage proportions, of the proposed confidence intervals are evaluated in a simulation study. For illustration, we consider applications to data from COVID-19 patients and from patients suffering from acute odontogenic infection. The proposed approach may also be adapted for constructing confidence intervals for other tree-based methods.
\end{abstract}
% Keywords 

\vspace{1ex}
\hrule width0.4\textwidth
\hspace{0.5ex}\footnotesize{\Letter\ \ Nikolai Spuck} 
\vspace{-0.5ex}

\vspace{-2.5ex}
\hspace{4.1ex}\footnotesize{spuck@imbie.uni-bonn.de}
\vspace{-1ex}
\normalsize

\section{Introduction}

Regression analysis is a powerful tool to quantify the association between an outcome variable and a set of covariates and to make inferences about the true parameter values. Classical statistical theory provides asymptotic properties for estimators of regression coefficients that allow performing hypothesis tests and constructing confidence intervals (CIs) based on their estimates. It is, however, a well known result that classical inference is invalid if the analysis involves a data-driven model selection procedure, that is, if the structure of a model's predictor function is determined in a data-driven way (e.g.~by forward or stepwise variable selection; \citealp{Taylor2015}), as statistical uncertainty induced by the model selection is neglected. 

%This issue is referred to as \textit{selective} or \textit{post selection inference} \citep{Berk2013, Fithian2014, Lee2016} and has been intensively studied in linear regression models \citep{Zhang2022}. 

%Consider data with an outcome variable $Y$ and $p$ covariates $\bs{X} = (X_1,\dots , X_p)$. Let  $\mathcal{D} = \{(y_i, \bs{x}_i)$, $i = 1, \dots, n\}$ be an independent sample of realizations of these random variables. Suppose that we want to make inference on a parameter $\beta$. The main principle of selective inference is to construct a $(1-\alpha)$ confidence interval $CI_{\beta}(\mathcal{D})$ that satisfies 
%\[
%\mathbb{P}(\beta \in CI_{\beta}(\mathcal{D}) \;|\; \beta \; \text{was selected}) \geq 1-\alpha\,,
%\]
%which means that the confidence interval conditions on the fact that fitting the regression model to $\mathcal{D}$ led to the selection of $\beta$. More recently, selective confidence intervals were proposed, among others, by Tibshirani et al. (2016) for sequential selection procedures, by Ruegamer \& Greven (2018) after likelihood-based model selection, and by Suzumura et al. (2017) for linear models including higher-order interactions.  

In this article we deal with this issue in the context of CIs for regression models with varying coefficients. This class of models first introduced by \citet{Hastie1993} generalizes the class of linear regression models, as they allow that coefficients of covariates change with the value of other variables, the so-called \textit{effect modifiers}. \citet{Fan2008} and \citet{park2015} gave comprehensive reviews on varying coefficient models and discussed several fitting approaches. In the past years varying coefficient models have been considered extensively, which has led to many extensions of the classical approach, see, for example, \citet{Wang2014}, \citet{Buergin2015, Buergin2017}, \citet{Lee2020}, and \citet{zhou2022}.  A large part of this methodology makes the basic assumption that the effect of each covariate is modified by a known set of potential effect modifiers, which is specified before model fitting. Then one determines the way the coefficients are modified. This prerequisite is relaxed by tree-structured varying coefficient (TSVC) models proposed by \citet{Berger2019}, which select the effect modifiers from the whole set of available covariates in a data-driven way and allow that the varying coefficients are caused by the interaction of several effect modifiers. The TSVC model applies recursive partitioning to identify relevant effect modifiers, which yields a separate tree $T_j$ for the linear effect of each covariate $X_j$, where each leaf node contains a partition-specific coefficient. Building a tree means to find a partition of the covariate space using binary splits, which induces a piecewise constant predictor function. In TSVC models, each split refers to a coefficient and is determined by an effect modifier and a corresponding splitting rule: For a metrically or ordinally scaled effect modifier~$X_k$, the splitting into two partitions has the form `$N_1 = N \cap \{X_k \leq c\}$ and $N_2 = N \cap \{X_k > c\}$', with regard to split point $c$, where $N$ denotes the parent node. 

While TSVC models may apply a concept of statistical significance to decide whether to split or not (described in detail in \citealp{Berger2019}), they so far do not allow for inference on the varying coefficients within the nodes of the fitted trees. Here, we fill this gap proposing a bootstrap-based approach. 

When constructing CIs for tree-structured varying coefficients one needs to take into account that the tree structures and therefore the partitions that differ in terms of their effect (defined by the effect modifiers and splitting rules) were estimated from the data (here denoted by $\mathcal{D}$). This is further aggravated by the fact that the trees suffer from high variability, that is, only a small change in the data $\mathcal{D}$ can lead to a rather different model structure (here denoted by $\mathcal{M}$), which may even strongly deviate from the true data generating process (DGP). Neglecting this uncertainty induced by the data-driven tree building procedure yields CIs that are likely to be too short. More specifically, it must be taken into account that the coefficients of interest $\beta_{jm}^{\mathcal{M}}$ (the linear effect of covariate $j$ in node $m$ of tree $T_j$) arises out of model structure $\mathcal{M}$, that is, that the model selection event $\widehat{\mathcal{M}} = \mathcal{M}$ occurred. 

The TSVC model~$\mathcal{M}$ considered in this paper is defined by a set of partitions $\mathcal{M} = \{\{ N_{jm},\, m = 1,...,M_{j}\},\, j = 1,...,p\}$.  Therefore, a $100(1-\alpha) \%$ CI of $\beta_{jm}^{\mathcal{M}}$ is supposed to satisfy
\begin{equation}
\label{TSVC_CI}
\mathbb{P}\left (\beta_{jm}^{\mathcal{M}} \in CI(\beta_{jm}^{\mathcal{M}}) \;|\; \widehat{\mathcal{M}} = {\mathcal{M}}  \right) \geq 1-\alpha\, ,
\end{equation}
which constitutes a so-called \textit{selective inference} or \textit{post selection inference} problem \citep{Berk2013, Fithian2014, Lee2016}. This issue has been intensively studied in linear regression models \citep{Zhang2022}. Selective CIs were proposed, among others, by \citet{tibshirani2016} for sequential variable selection procedures, by \citet{rugamer2018} after likelihood-based model selection in linear models, and by \citet{suzumura2017} for linear models including higher-order interactions.  More recently, \citet{Zhao2022} proposed a selective inference approach for LASSO-based varying coefficient models, \citet{Ruegamer2022} investigated selective inference for additive and linear mixed effect models, and \citet{Zrnic2023} address the selective inference problem by building on the framework of algorithmic stability. With respect to tree-structured models, \citet{Gottard2023} proposed a simple splitting of the data into training data for fitting the model and test data for conducting inference, which, however, comes at the price that only a subset of the observations is used for tree building and for subsequent inference. \citet{Loh2019} proposed bootstrap-calibrated CIs within the GUIDE regression tree framework and \citet{Neufeld2022} proposed ``Tree-Values'', a selective inference framework for regression trees.

\begin{sloppypar}
In line with the principle of selective inference, we propose a parametric bootstrap approach to construct $100(1-\alpha)\%$ percentile CIs $CI_P (\beta_{jm}^{\mathcal{M}})$ for a TSVC model $\mathcal{M}$ with varying coefficients $\beta_{jm}^{\mathcal{M}}\,,j=1,\hdots,p\,,m=1,\hdots,M_j$, that satisfy Equation~\eqref{TSVC_CI}. 
\end{sloppypar}

The remainder of this paper is organized as follows: In Section~2, the class of TSVC models and the fitting procedure are described. Section~3 outlines our parametric bootstrap approach for constructing percentile CIs of the varying coefficients. To assess coverage proportions of the proposed CIs, we conducted a simulation study presented in Section~4. In the simulation study, we also contrast our proposal to bootstrap-calibrated CIs. In Section~5, we show the results of two applications fitting TSVC models to data of patients suffering from COVID-19 and acute odontogenic infection, respectively. Finally, our findings are summarized and discussed in Section~6.

\section{Tree-structured varying coefficients}

Let $Y$ be a outcome variable of interest and $\bs{X} =( X_1,\dots , X_p)$ be explanatory variables that are ordinally or metrically scaled, or dummy-coded representations of nominal variables.
In generalized regression models it is assumed that the outcome $Y$ given the values of covariates $\bs{X}$ follows a distribution from the exponential family. The expected outcome is related to the covariate vector in the form $\E (Y|\bs{X}) = g^{-1}(\eta (\bs{X}))$, where $g(\cdot )$ denotes a suitable link function and $\eta(\cdot )$ denotes the predictor function. Most frequently it is assumed that the predictor function is characterized by a linear combination of the covariates. In the more general varying coefficient model introduced in the seminal work by \citet{Hastie1993}, the predictor function is given by
\begin{equation}
\label{vc}
\eta(\bs{X}, \bs{Z}) = \beta_0 + \beta_{1}(Z_{1})X_{1} + \dots + \beta_{p}(Z_{p})X_{p}\, ,
\end{equation}
where $Z_1,\dots, Z_p$ denote (additional) random variables that serve as \textit{effect modifiers} and change the linear effects of $X_{1},\dots , X_{p}$ through unspecified functional forms. 

The model with predictor \eqref{vc} requires the effect modifiers to be specified beforehand. In practice, however, it is often unclear which variable modifies the effect of another variable. In addition, each varying coefficient may not be determined by just one variable and the effect may be driven by an interaction between several effect modifiers. To address these issues, \citet{Berger2019} proposed the tree-structured varying coefficient (TSVC) model, which applies a recursive partitioning method to detect relevant effect modifiers. Since the effect modifiers are inherently selected by the tree building algorithm, only the set of covariates $X_1,\dots, X_p$ is considered for modeling. If effect modifiers are present, they are from this set and modify coefficients of covariates from this set. The predictor function of a TSVC model $\mathcal{M}$ is given by
\begin{equation}
\eta^{\mathcal{M}} (\bs{X}) = \beta_0^{\mathcal{M}} + \beta_1^{\mathcal{M}}(\bs{X}_{[-1]})X_{1} + \dots + \beta_p^{\mathcal{M}}(\bs{X}_{[-p]})X_{p}\, ,
\end{equation}
 where $\bs{X}_{[-j]}$ denotes the set of covariates $X_1,\dots, X_p$ excluding $X_{j}$. By definition, the effect of each covariate can be modified by each other covariate except itself. The functions $\beta_j^{\mathcal{M}}(\cdot)$ are each determined by a tree structure. This means that each function $\beta_{j}^{\mathcal{M}}(\cdot)$ sequentially partitions the observations into disjoint subsets $N_{jm},\, m = 1,..., M_j$, based on the values of the selected effect modifiers and assigns a different regression coefficient for $X_j$ to each partition $N_{jm}$. These functions can be written as
\begin{equation}
\label{TSVC}
\beta_j^{\mathcal{M}}(\bs{X}_{[-j]}) = \sum_{m = 1}^{M_j} \beta_{jm}^{\mathcal{M}}\,I(\bs{X}_{[-j]}\in N_{jm})\, ,
\end{equation}
where $I(\cdot)$ denotes the indicator function. Hence, the structure of a TSVC model~$\mathcal{M}$ is characterized by the set of partitions $\mathcal{M} = \{\{N_{jm}\,, m = 1,\dots, M_j\}\,, j = 1,\dots, p\}$. Each coefficient is derived from binary splits successively partitioning the observations of one parental node into two child nodes (cf.\@ \citealp{Hastie2009}). We start from a model with non-varying linear effects, only. Then, the first split yields model $\mathcal{M}^{[1]}$  with predictor
\begin{align*}
\label{step1}
\eta^{\mathcal{M}^{[1]}} (\bs{X}_i ) = &\, \beta_0^{\mathcal{M}^{[1]}} + \beta_{1}^{\mathcal{M}^{[1]}}X_{1} + \dots +\left(\beta_{j1}^{\mathcal{M}^{[1]}}I(X_{k}\leq c_k) + \beta_{j2}^{\mathcal{M}^{[1]}}I(X_{k}> c_k)\right)X_{j} \nonumber \\
& + \dots + \beta_p^{\mathcal{M}^{[1]}} X_{p}\, ,
\end{align*}
where $c_k$ is the split point in effect modifier $X_k$ selected by the algorithm regarding the effect of $X_j$, $\beta_{j1}^{\mathcal{M}^{[1]}}$ is the linear effect of $X_j$ in partition $\{X_{k} \leq c_k\}$ and $\beta_{j2}^{\mathcal{M}^{[1]}}$ is the linear effect of $X_{j}$ in partition $\{X_{k}>c_k\}$ adjusted for the other effects in ${\mathcal{M}^{[1]}}$.  Hence, after the first step, the varying coefficient of $X_j$ is determined by $\beta_j^{\mathcal{M}^{[1]}}(X_{k}) = \beta_{j1}^{\mathcal{M}^{[1]}}I(X_{k}\leq c_k) + \beta_{j2}^{\mathcal{M}^{[1]}}I(X_{k}> c_k)$. In the next step, either a different coefficient is selected for splitting or the same coefficient is further modified. If the coefficient of variable $X_\ell$ is split in $X_{r}$ at split point $c_r$ this yields the predictor
\begin{align*}
\eta^{\mathcal{M}^{[2]}} (\bs{X} ) =&\, \beta_0^{\mathcal{M}^{[2]}} + \beta_{1}^{\mathcal{M}^{[2]}}X_{1} + \dots +\left(\beta_{j1}^{\mathcal{M}^{[2]}}I(X_{k}\leq c_k) + \beta_{j2}^{\mathcal{M}^{[2]}}I(X_{k}> c_k)\right)X_j   \nonumber \\
 & +  \left(\beta_{\ell1}^{\mathcal{M}^{[2]}}I(X_{r}\leq c_r) + \beta_{\ell2}^{\mathcal{M}^{[2]}}I(X_{r}>c_r)\right)X_{\ell} +\dots + \beta_p^{\mathcal{M}^{[2]}} X_{p}\, ,
\end{align*} 
where $\beta_{\ell1}^{\mathcal{M}^{[2]}}$ denotes the effect of $X_\ell$ in $\{X_{r}\leq c_r\}$ and $\beta_{\ell2}^{\mathcal{M}^{[2]}}$ denotes the effect of $X_{\ell}$ in $\{X_{r}>c_r\}$. That is, the varying coefficient of $X_\ell$ has the form $\beta_{\ell}^{\mathcal{M}^{[2]}}(X_{r}) = \beta_{\ell1}^{\mathcal{M}^{[2]}}I(X_{r}\leq c_r) + \beta_{\ell2}^{\mathcal{M}^{[2]}}I(X_{r}>c_r)$. Further splits are performed analogously until a predefined stopping criterion is met (see below for details). In each step a so far non-varying effect turns into a varying coefficient or an already selected varying coefficient is split once more. 

\subsection*{Sketch of the fitting procedure}
\label{fit}
In each step of the tree building algorithm, the best splitting rule from among all possible combinations of covariate $X_j$, respective candidate effect modifier $X_k, k \neq j$, and split point is selected, starting from a linear predictor without varying coefficients. For this, all candidate models with one additional split are evaluated and the best-performing one that yields the smallest deviance is selected. In generalized regression models the deviance is a quite natural measure of the model fit. This criterion is also equivalent to minimizing the entropy, which has been used as a splitting criterion already in the early days of tree construction \citep{Breiman1984}. Note that, in contrast to common trees, in each step of the algorithm all the observations are used to derive new estimates of the model parameters. This ensures that one obtains valid estimates of the
different components together with the splitting rule.

To determine the optimal number of splits and hence the size of the trees, \citet{Berger2019} proposed an early stopping strategy based on permutation tests. This offers an approximate solution to control the global type I error rate (that is, the proportion of falsely identified covariates with varying coefficients). Here, we use an alternative post-pruning strategy, where a large model is grown first and is then pruned to an adequate size to avoid overfitting. Running the stepwise TSVC algorithm (with a sufficiently large number of splits) yields a sequence of nested models that are assessed with regard their goodness of fit using a likelihood-based criterion. Subsequently, the best-performing model is selected and fitted on the whole data with the corresponding number of splits. We suggest to select the optimal number of splits by minimizing the Bayesian information criterion (BIC; \citealp{Schwarz1978}). The BIC of a TSVC model $\mathcal{M}^{[s]}$ is given by
\begin{equation}
BIC\left (\mathcal{M}^{[s]} \right ) = - 2\ln \left ( L_{\mathcal{M}^{[s]}} \right ) + s \log (n)\, ,
\end{equation}
where $ L_{\mathcal{M}^{[s]}}$ denotes the maximized value of the likelihood function of model $\mathcal{M}^{[s]}$, $s = \sum_{j=1}^{p}{(M_j-1)}$ is the total number of performed splits, and $n$ denotes the number of observations.

\section{Selective confidence intervals}

To demonstrate the selective inference problem formalized in Equation~\eqref{TSVC_CI} that comes with the TSVC model, we consider a simple example with two metrically scaled covariates $\bs{X} = (X_1, X_2)$. Let $Y$ be the outcome variable that follows a normal distribution, $Y \sim N \left(\mu (\bs{X}), \sigma^2\right)$, with conditional expectation
\begin{equation}
\label{exDist}
 \mu (\bs{X}) = \beta_0+\beta_{11}\,I(X_{2}\leq c_2)\,X_1 + \beta_{12}\,I(X_{2}> c_2)\,X_{1} + \beta_{2}\,X_{2}\, .
\end{equation}
Hence, there exists a varying linear effect of $X_1$ on $Y$ modified by $X_2$. Specifically, two regions with different linear effects of $X_1$ are present: region $R_{11}= \{ X_2 \leq c_2\} $ with linear effect $\beta_{11}$ and region $R_{12} = \{ X_2 > c_2\}$ with linear effect $\beta_{12}$. The linear effect of $X_2$ remains the same across the whole covariate space, that is, $R_{2} = \{X_1\in \mathbb{R}\}$.

\begin{sloppypar}
Assume a TSVC model $\mathcal{M}_{1}$ is fitted to a sample  $\mathcal{D}_1 = \{(y_i^{(1)}, \bs{x}^{(1)}_i=(x_{i1}^{(1)}, x_{i2}^{(1)})), i = 1,\dots, n\}$, where the values of the outcome variable $y_i^{(1)}$ were drawn from the normal distribution with conditional expectation~\eqref{exDist}. If the structure of the fitted model coincides with the true DGP (one split of $X_2$ in the linear effect of $X_1$ with split point $c_2$), that is, $\widehat{\mathcal{M}}=\mathcal{M}_1 = \{\{R_{11}, R_{12} \}, \{ R_2 \} \}$, one obtains estimates $\hat{\beta}_{11}$, $\hat{\beta}_{12}$, $\hat{\beta}_{2}$ of the underlying DGP parameters.  However, since the structure of model $\mathcal{M}_{1}$ is determined based on the sample $\mathcal{D}_1$, the partitions detected by the algorithm are likely to deviate from the true underlying regions, which means $\mathcal{M}_1 \neq \{\{R_{11}, R_{22}\}, \{R_{2}\} \}$. For instance, assume the predictor function of model~$\mathcal{M}_1$ is given by
\end{sloppypar}
\begin{equation*}
\eta^{\mathcal{M}_{1}}(\bs{X}) = \hat{\beta}_0^{\mathcal{M}_1} + \hat{\beta}_{11}^{\mathcal{M}_1}\,X_1 + \hat{\beta}_{21}^{\mathcal{M}_1}\,I(X_1\leq c_1)\,X_2 + \hat{\beta}_{22}^{\mathcal{M}_1}\,I(X_1> c_1)\,X_2\; .
\end{equation*}
In this model, the interaction between $X_1$ and $X_2$ was found correctly but its form is flawed. Specifically, we have that $\mathcal{M}_{1} = \{\{N_{11}\}, \{N_{21}, N_{22}\}\}$ with $N_{11} = \{X_2 \in \mathbb{R} \}, N_{21} = \{X_1 \leq c_1\}$ and $N_{22} = \{ X_1 > c_1\}$. As the structure of the model is misspecified, none of the estimated coefficients $\bs{\hat{\beta}}^{\mathcal{M}_1} = (\hat{\beta}_{11}^{\mathcal{M}_1}, \hat{\beta}_{21}^{\mathcal{M}_1}, \hat{\beta}_{22}^{\mathcal{M}_1})$ matches the true underlying effects $\bs{\beta} = (\beta_{11}, \beta_{12},\beta_2)$ from the DGP~\eqref{exDist}. Instead, as outlined analogously for data-driven variable selection in linear models by \citet{Berk2013}, the coefficient vector $\bs{\hat{\beta}}^{\mathcal{M}_1}$ actually is an estimate to the solution of the optimization problem 
\begin{align}\label{eq:optim}
\bs{\beta}^{\mathcal{M}_1} = \E\big(\bs{\hat{\beta}}^{\mathcal{M}_1}\big)  &= \argmax_{\bs{b^{\mathcal{M}_1}}} \, \E \left(L_{\mathcal{M}_{1}}\big(\bs{b}^{\mathcal{M}_1}\,|\; Y, \bs{x}^{(1)}\big)\right)\nonumber \\ 
&  = \argmax_{\bs{b^{\mathcal{M}_1}}} \,L_{\mathcal{M}_{1}}\big(\bs{b}^{\mathcal{M}_1}\,|\; \mu(\bs{x}^{(1)}), \bs{x}^{(1)}\big)\; ,
\end{align}
where $L_{\mathcal{M}_1}(\cdot)$ denotes the likelihood function of model $\mathcal{M}_{1}$. Note that, following \citet{Berk2013}, the expectation is evaluated only with regard to the outcome variable $Y$, and  the values of the covariates are treated as fixed at the observed values $\bs{x}^{(1)} = (\bs{x}^{(1)}_1, \bs{x}^{(1)}_2)$. By definition, $\bs{\beta}^{\mathcal{M}_1}$ maximizes the likelihood of the model with the fixed structure $\mathcal{M}_{1}$ using the expected values of the outcome variable given the observed values of the covariates (i.e.\@ assuming that the DGP is known). In addition, Equation~\eqref{eq:optim} implies that the target effects $\beta_{jm}^{\mathcal{M}}$ in different models are generally different.
For example, applying the TSVC fitting procedure to another sample $\mathcal{D}_{2}=  \{(y^{(2)}_i, \bs{x}^{(2)}_i=(x_{i1}^{(2)}, x_{i2}^{(2)})), i = 1,...,n\}$ yields a new model $\mathcal{M}_2$ with a potentially different structure. Assume the predictor function of model~$\mathcal{M}_2$ is given by
\begin{align*}
\eta^{\mathcal{M}_{2}}(\bs{X}) = \hat{\beta}_0^{\mathcal{M}_2} &+ \hat{\beta}_{11}^{\mathcal{M}_2}\,I(X_2\leq c_2)\,X_1 + \hat{\beta}_{12}^{\mathcal{M}_2}\,I(X_2\leq c_2)\,X_1  \nonumber \\ 
& + \hat{\beta}_{21}^{\mathcal{M}_2}\,I(X_1\leq c_1\,)X_2 + \hat{\beta}_{22}^{\mathcal{M}_2}\,I(X_1> c_1)\,X_2\; .
\end{align*}
Even though the estimated coefficients $\hat{\beta}_{21}^{\mathcal{M}_1}$ and $\hat{\beta}_{21}^{\mathcal{M}_2}$ of $X_2$ refer to same partition~$N_{21}$, they do not estimate the same effect, because model structures $\mathcal{M}_1$ and $\mathcal{M}_2$ do not match and were adjusted for differently structured effects of $X_1$. 

%Importantly, if the structure of the data generating process was identified correctly, that is $\widehat{\mathcal{M}} = \{\{S_{11}, S_{12} \}, \{ S_2 \} \}$, maximizing the expected likelihood function yields the true coefficients $\bs{\beta}$ from the data generating process in Equation~\eqref{exDist}.

More generally, given a data set $\mathcal{D} = \{(y_{i}, \bs{x}_{i} = (x_{i1},...,x_{ip}), i = 1,...,n\}$ and a TSVC model with selected structure $\mathcal{M} = \{\{ N_{jm}\,, m = 1,...,M_j\}\,, j = 1,...,p\}$ fitted to the data, the corresponding coefficient vector $\bs{\hat{\beta}^{\mathcal{M}}}$ estimates the \textit{best approximating varying linear coefficients} given the selected model structure $\mathcal{M}$ defined by
 \begin{equation}
\label{defBetaM}
\bs{\beta^{\mathcal{M}}} = \E\big(\bs{\hat{\beta}}^{\mathcal{M}}\big) = \argmax_{\bs{b^{\mathcal{M}}}} \,L_{\mathcal{M}}\big(\bs{b}^{\mathcal{M}}\,|\; \mu(\bs{x}), \bs{x}\big)\; .
\end{equation}
Our objective is to construct selective CIs for the coefficients $\beta_{jm}^{\mathcal{M}}$ as defined in Equation~\eqref{defBetaM} that satisfy Equation~\eqref{TSVC_CI}. For this purpose, the distribution of~$\bs{y}$ conditional on the model selection event $\widehat{\mathcal{M}} = \mathcal{M}$ needs to be considered. Note that if the structure of the selected model coincides with the structure of the DGP, the best approximating varying linear coefficients match the true effects of the DGP. In linear regression models with LASSO penalization \citet{Lee2016} found that if the selection event $\widehat{\mathcal{M}} = \mathcal{M}$ can be characterized by a set of inequalities $\bs{Ay} \leq \bs{b}$, where $\bs{A}$ and $\bs{b}$ must not depend on $\bs{y}$, $\widehat{\mathcal{M}} = \mathcal{M}$ constitutes a \textit{linear selection event} and exact statistical inference of the coefficients conditional on the selection event can be performed. Specification of the selection event for TSVC models, however, would require a vast number of inequalities. The main reason is that the TSVC algorithm involves the fitting of several trees, which is considerably more complex than fitting of a single tree or a predictor function with interactions of predefined order (scenarios investigated by \citealp{Neufeld2022} and \citealp{suzumura2017}, respectively). 
%Due to the flexible applicability of the TSVC fitting procedure in generalized regression settings, the TSVC model selection event may hardly be formulated as a linear selection event and exact statistical inference cannot be conducted. While \citet{Taylor2018} proposed an asymptotic approach to address nonlinear selection events in the context of generalized regression for penalized likelihood models, no method for dealing with the complexity of the TSVC fitting procedure is currently available. The main issue is that the definition of the TSVC model selection event $\widehat{\mathcal{M}} = \mathcal{M}$ requires a large number of inequalities. 
Specifically, in the first iteration of the TSVC algorithm, the event of selecting one splitting rule is characterized by $p(p-1)n$ inequalities, assuming $p$ continuous covariates with $n$ possible split points each. Each inequality specifies that the maximal likelihood value of a model that results from one of the possible splitting rules (i.e. a combition of covariate $X_j$, effect modifier $X_k, k\neq j$, and split point) is lower than the maximal likelihood value of the model with the selected splitting rule. Overall, $\mathcal{O}(np^2S)$ inequalities are required to describe the selection of one particular sequence of nested TSVC models $\mathcal{M}^{[s]}, s =1,\dots, S$, and an optimal model $\mathcal{M}$ out of it.  
Since there are cases where the same model structure $\mathcal{M}$  can arise from a number of different sequences of nested models (e.g.\@ when the same splits are performed in a different order), the conditioning set that characterizes the selection event $\widehat{\mathcal{M}} = \mathcal{M}$ is a union of sets defined by these inequalities. 
%Moreover, after the selected number of splits is reached, any combination of additional splitting rules can result in the same model structure as long as the respective models' BIC is larger than the BIC of the selected model. 

To tackle this complex mechanism, we propose a parametric bootstrap approach tailored to the selective inference problem at hand (described in detail in Sections~\ref{subsec:boot} and~\ref{subsec:param}). Basically, given a TSVC model $\mathcal{M}$ fitted to data $\mathcal{D}$ with coefficients $\beta^{\mathcal{M}}_{jm}$, we~(i) generate samples $\mathcal{D}_b$ by drawing new values of the outcome variable $Y$ while keeping the covariate values fixed using a parametric bootstrap scheme and (ii) calculate estimates of $\beta^{\mathcal{M}}_{jm}$ from the TSVC model $\mathcal{M}_b$ fitted to $\mathcal{D}_b$ in order to construct percentile CIs based on these estimates.

\subsection{Calculating bootstrap effect estimates for given $\mathcal{M}$}\label{subsec:boot}

To construct a CI for the coefficient $\beta_{jm}^{\mathcal{M}}$ from a given model with structure $\mathcal{M}$ satisfying Equation~\eqref{TSVC_CI}, we compute estimates for $\beta_{jm}^{\mathcal{M}}$ from a set of bootstrap samples $\mathcal{D}_b,\, b = 1,\hdots,B$ (see Section~\ref{subsec:param} for details on the applied bootstrap sampling scheme). A naive approach, which simply enforces the structure of the original model $\mathcal{M}$ on each sample, would imply that the model structure is predefined and neglect the uncertainty induced by the data-driven tree building procedure. To account for this uncertainty, we first apply the TSVC fitting procedure to the samples~$\mathcal{D}_b$, resulting in $B$ different models $\mathcal{M}_b$ of potentially different form. For each $b$ the predictor function of model $\mathcal{M}_b$ is given by
\begin{equation*}
\label{bootModel1}
\eta^{\mathcal{M}_b}(\textbf{\textit{X}}) = \hat{\beta}^{\mathcal{M}_b}_0 + \hat{\beta}_1^{\mathcal{M}_b}(\textbf{\textit{X}}_{[-1]})X_{1} + \hdots + \hat{\beta}_p^{\mathcal{M}_b}(\textbf{\textit{X}}_{[-p]})X_{p}\,
\end{equation*}
with
\begin{equation*}
\label{bootModel2}
\hat{\beta}_j^{\mathcal{M}_b} (\bs{X}_{[-j]}) = \sum_{m = 1}^{M_j^{(b)}} \hat{\beta}_{jm}^{\mathcal{M}_b}\,I(\bs{X}_{[-j]}\in N^{(b)}_{jm}) \, .
\end{equation*}
Secondly, we determine an estimate of the coefficient of interest $\beta^{\mathcal{M}}_{jm}$ from the original model based on bootstrap model $\mathcal{M}_b$ by averaging  the node-specific effect estimates $\hat{\beta}_{jm}^{\mathcal{M}_b}$ with regard to the corresponding partition $N_{jm}$ from the original model yielding
\begin{equation}
\label{bootEst}
\bar{\beta}_{jm}^{(b)} = \frac{1}{|N_{jm}|}\sum_{i:\textbf{\textit{x}}_i \in N_{jm}} \hat{\beta}_{j}^{\mathcal{M}_b}(\textbf{\textit{x}}_{i[-j]})\, .
\end{equation}
This means, for each covariate $X_j$ each observation is assigned to one of the subsets $N_{jm}$ that was identified by the original model $\mathcal{M}$, and subsequently the average value of the function $\hat{\beta}_{j}^{\mathcal{M}_b}(\cdot)$  from model $\mathcal{M}_b$ across the observations in $N_{jm}$ is calculated. Therefore, Equation~\eqref{bootEst} defines an estimate of $\beta_{jm}^{\mathcal{M}}$ for bootstrap sample $\mathcal{D}_b$ that accounts for the uncertainty induced by the data-driven tree building. 

Finally, a $100(1-\alpha)\%$ percentile CI for $\beta_{jm}^{\mathcal{M}}$ is constructed as 
\begin{equation}
\label{perc}
CI_{P}\left (\beta_{jm}^{\mathcal{M}}\right) = \left[\bar{\beta}_{jm}^{\alpha/2}, \bar{\beta}_{jm}^{1-\alpha/2}\right], 
\end{equation} 
\begin{sloppypar}
where $\bar{\beta}_{jm}^{q}$ denotes the $100q$-th percentile of the set of bootstrap estimates $\bar{\beta}_{jm}^{(1)}, \dots , \bar{\beta}_{jm}^{(B)}$.
\end{sloppypar}

\subsection{Parametric bootstrap procedure}\label{subsec:param}

By the definition in Equation~\eqref{bootEst} one can determine bootstrap estimates of the coefficients of interest $\beta_{jm}^{\mathcal{M}}$. Yet, calculating these estimates does in itself not condition on the model selection event $\widehat{\mathcal{M}} = \mathcal{M}$. To take this into account, we mimic the conditioning by applying a parametric bootstrap scheme, which is based on the original model $\mathcal{M}$. 

For each observation~$i$, the value $y_i^{(b)}$ in bootstrap sample $\mathcal{D}_b$ is drawn from the conditional distribution of $Y|\, \bs{X} = \bs{x}_i$ given the fitted TSVC model $\mathcal{M}$, following the parametric bootstrap sampling scheme described in \citet{Efron1993}. That is, the new outcome values $y_{i}^{(b)}$ are generated from a distribution with expectation
\begin{equation}
\label{paramboot}
\E (Y|\, \bs{X} = \bs{x}_i) = g^{-1}\left (\eta^{\mathcal{M}}(\bs{x}_i)\right ) \, ,
\end{equation}
where $\eta^{\mathcal{M}}(\cdot)$ is the predictor function of the original TSVC model fitted to $\mathcal{D}$. Of note, we keep the values of the covariates $\bs{X}$ fixed at the observed values~$\bs{x}_i$, which is in line with the definition of the best approximating varying linear coefficients in Equation~\eqref{defBetaM}. 

If a Gaussian TSVC model is considered, $g(\cdot)$ is the identity link and the new outcome values of the bootstrap samples are drawn using
\begin{equation*}
y_{i}^{(b)} \sim N \left ( \eta^{\mathcal{M}}(\bs{x}_i),\, \hat{\sigma}^2_{\varepsilon}\right),
\end{equation*}
where $N(\cdot,\cdot )$ denotes the normal distribution and $\hat{\sigma}^2_{\varepsilon}$ is the residual variance of model $\mathcal{M}$. For a binary logistic TSVC model, the new outcome values are generated as
\begin{equation*}
y_{i}^{(b)} \sim \text{Bin}\left(1,\,\text{logit}\left (\eta^{\mathcal{M}}(\bs{x}_i) \right ) \right )\, ,
\end{equation*}
where $\text{Bin}(\cdot, \cdot)$ denotes the binomial distribution. In general, this approaches allows to generate $B$ bootstrap samples $\mathcal{D}_b = \{(y_{i}^{(b)}, \bs{x}_i),\, i = 1,\dots n\}$ for any kind of generalized TSVC model. 
%Since the parametric bootstrap is performed based on the original model $\mathcal{M}$, TSVC models fitted to the generated samples $\mathcal{D}_b$ are more likely to exhibit the same or a structure similar to the original model, thereby mimicking the conditioning on the model selection event $\widehat{\mathcal{M}} = \mathcal{M}$.

To summarize, given a TSVC model $\mathcal{M}$ fitted to data $\mathcal{D}$, we propose to perform the following steps to construct $100(1-\alpha)\%$ CIs for the coefficients $\beta_{jm}^{\mathcal{M}}$:
\begin{enumerate}
\item[1.] \textbf{Bootstrap sampling:} Generate $B$ bootstrap samples $\mathcal{D}_b = \{(y_{i}^{(b)}, \bs{x}_i), i = 1,\dots n\}$ by sampling new outcome values from the conditional distribution of the outcome variable with expectation~\eqref{paramboot}.
\item[2.] \textbf{Model fitting:} Apply the TSVC fitting procedure described in Section~\ref{fit} to each sample $\mathcal{D}_b$ in order to obtain a model $\mathcal{M}_b$ for each sample.  
\item[3.] \textbf{Calculating bootstrap estimates:} Determine the estimates $\bar{\beta}_{jm}^{(b)}$ as defined in Equation~\eqref{bootEst} for each model $\mathcal{M}_b$.
\item[4.] \sloppypar{\textbf{Percentile intervals:} Construct percentile CIs by computing the $100(\alpha/2)$-th and $100(1-\alpha/2)$-th percentiles of the set of bootstrap estimates $\bar{\beta}_{jm}^{(1)}, \dots , \bar{\beta}_{jm}^{(B)}$ as described in Equation~\eqref{perc}. }
\end{enumerate}

\section{Simulation study}

To assess coverage proportions of the proposed parametric bootstrap percentile CIs, we considered different simulation scenarios.  The aims of the simulation study were (i) to evaluate how the coverage of the proposed CIs is affected by the structure of the DGP, (ii) to investigate the effect of sample size and noise in the DGP on the coverage proportions, and (iii) to compare coverage proportions of the proposed CIs to those based on alternative types of CIs (e.g. simple asymptotic normal distribution-based Wald intervals). The scenarios were based on a linear DGP without varying effects (scenario 1), a tree-structured varying effect DGP (scenario 2), and a tree-structured varying effect DGP where effect modifiers were prespecified before model fitting (scenario 3). Further details on the DGPs will be given in the following subsections.
In each replication of the three scenarios, a TSVC model was fitted to the data, where the maximal number of splits was set to $S = 5$ and the BIC was used to determine the optimal number of splits. For the resulting coefficients of interest, $90\%$ as well as $95\%$ CIs were constructed using the following methods:
\begin{itemize}
\item[(i)] simple asymptotic normal distribution-based Wald type CIs (\textit{Wald}),
\item[(ii)] our proposed parametric bootstrap percentile CIs (\textit{Parametric percentile}), and
\item[(iii)] Wald type CIs, where an adjusted $\alpha$-level is determined via bootstrap calibration to account for the uncertainty induced by the tree building (\textit{Bootstrap calibration}; \citealp{Loh2019}).
\end{itemize}
The $100(1-\alpha)\%$ Wald type CIs are calculated as
\begin{equation}
CI_{W}\left(\beta_{jm}^{\mathcal{M}}\right) = \left[\hat{\beta}_{jm}^{\mathcal{M}} + z_{\alpha/2}\text{SE}(\hat{\beta}_{jm}^{\mathcal{M}}),\, \hat{\beta}_{jm}^{\mathcal{M}} + z_{1-\alpha/2}\text{SE}(\hat{\beta}_{jm}^{\mathcal{M}})\right]\, ,
\end{equation}
where $z_{q}$ denotes the $100q$-th percentile of the standard normal distribution and SE($\hat{\beta}_{jm}^{\mathcal{M}}$) denotes the standard error of the coefficient estimate (not including the uncertainty induced by the model selection).
The bootstrap calibration method introduced by \citet{Loh2019} is designed to construct CIs for coefficients of regression models that were fitted on subgroups identified by a GUIDE regression tree. We applied a version of their algorithm adapted to TSVC models. For a detailed description of the approach, see the Supplementary Material. 
The proposed parametric percentile CIs and the bootstrap-calibrated CIs were constructed based on $B=1000$ bootstrap samples, respectively. 

\begin{sloppypar}
In the three simulation scenarios we considered a continuous outcome variable and either two or three covariates with a potential effect on the outcome, where $X_1, X_2\sim N(0,1)$ and $X_3\sim \text{Bin}(1, 0.5)$. We considered sample sizes of ${n\in \{200, 500, 1000\}}$ and DGPs with normally distributed error terms with standard deviations of $\sigma_{\varepsilon} \in \{1,2\}$. 
\end{sloppypar}

Coverage proportions were calculated based on $R=5000$ replications. For the varying linear coefficients of a covariate $X_j$, the average coverage proportion was calculated as
\begin{equation*}
C_j = \frac{1}{R}\sum_{r = 1}^{R}\frac{1}{M_j^{r}}\sum_{m = 1}^{M_j^{r}} I\left(\beta_{jm}^{\mathcal{M}_r} \in CI(\beta_{jm}^{\mathcal{M}_r})\right)\, ,
\end{equation*}
where $\mathcal{M}_r$ denotes the TSVC model fitted in the $r$-th replication and $M_{j}^r$ denotes the number of coefficients of $X_j$ in model $\mathcal{M}_r$. The average coverage proportion across all covariates is then given by 
\begin{equation*}
C_{\text{av}} = \frac{1}{p} \sum_{j = 1}^{p} C_j\, .
\end{equation*}

\subsection{Linear DGP}

In the first scenario,  $X_1$ and $X_2$ were included as covariates and potential effect modifiers in the TSVC fitting procedure. The DGP of the first scenario was given by
\begin{equation}
y_{i} = 0.25\,x_{i1} + \varepsilon_i\, ,\,  i = 1,\dots , n\, ,
\end{equation}
which means that $X_1$ has a simple non-varying linear effect and $X_2$ is non informative. The proportions of variance explained by $X_1$ were approximately $0.06$ ($\sigma_{\varepsilon} = 1$) and $0.02$ ($\sigma_{\varepsilon} = 2$).

\begin{table*}[!t]
\caption{Average number of splits based on 5000 replications when fitting a TSVC model for a linear DGP (scenario 1). The first row shows the average total number of splits per replication in the TSVC models. Below the average number of splits per replication is reported separately for each combination of covariate and effect modifier. The column \textit{DGP} contains the true number of splits.}
\label{sc1Splits}
\begin{center}
\begin{footnotesize}
\begin{tabularx}{\textwidth}{l l l l Y Y Y Y Y Y Y Y}
\toprule
Covariate & Effect modifier & DGP & $n$ & \multicolumn{2}{c}{200} & \multicolumn{2}{c}{500} & \multicolumn{2}{c}{1000} \\
  & & & $\sigma_\varepsilon$&1 & 2 & 1 & 2  &1& 2\\ \hline
 --- & --- & \textbf{0} & & \textbf{0.63} & \textbf{0.63} & \textbf{0.34} & \textbf{0.34} & \textbf{0.22} & \textbf{0.22} \\
$X_1$ & $X_2$ & 0 & & 0.32 & 0.32 & 0.18 & 0.18 & 0.11 & 0.11 \\
$X_2$ & $X_1$ & 0 & & 0.31 & 0.31 & 0.17 & 0.17 & 0.11 & 0.11 \\
\bottomrule
\end{tabularx}	
\end{footnotesize}
\end{center}		
\end{table*}

\begin{figure*}[!t]
\centering
\includegraphics[width = 11.9cm]{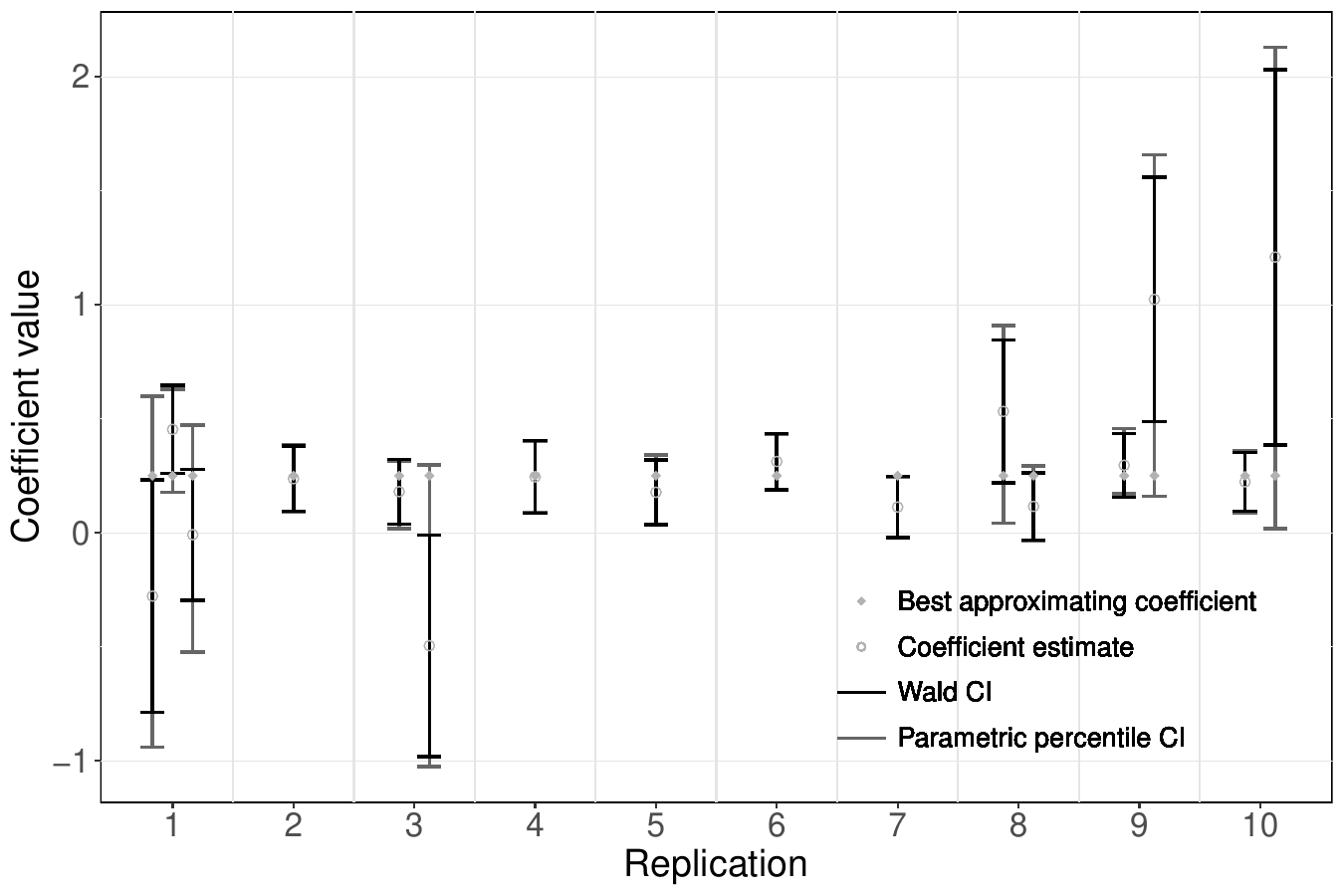}
\caption{Fitted tree-structured varying coefficients and 95\% CIs for a linear DGP (scenario 1). The figure shows the estimated effects $\hat{\beta}^{\mathcal{M}_r}_{jm}$ of the original model fitted to the data $\mathcal{D}$, the best approximating linear coefficients $\beta^{\mathcal{M}_r}_{jm}$ as defined in Equation~\eqref{defBetaM}, corresponding 95\% Wald type and parametric percentile CIs for the varying linear coefficients of $X_1$ for 10 exemplary replications. The number of coefficients of $X_1$ in the true linear DGP is $1$. The underlying data were drawn from a linear DGP with $n = 200$ and $\sigma_{\varepsilon} = 1$.}
\label{CoverageSC1}
\end{figure*}

Table~\ref{sc1Splits} shows that the number of splits performed by the TSVC fitting procedure increased with lower sample size but appeared to be unaffected by the standard deviation of the error term. Hence, the structure of the model tended to be closer to the DGP with larger sample sizes. Of note, the average number of falsely performed splits was nearly equal for the informative covariate $X_1$ and the non-informative covariate~$X_2$.  

Exemplary results of the fitted CIs are depicted in Figure~\ref{CoverageSC1}. The figure illustrates that the proposed parametric percentile CIs coincided with the simple Wald type CIs in cases where no splits were performed (i.e. if there was only one coefficient of~$X_1$, as, for example, in replication 2). In cases with more than one coefficient (for example in replication 1), the parametric percentile CIs were wider and were therefore more likely to cover the best approximating coefficient. From Figure~\ref{CoverageSC1} it is also seen, that unlike the Wald type CIs, the parametric percentile CIs were not necessarily symmetric around the coefficient estimate. Note that, due to the linear DGP without varying coefficients, the best approximating coefficient equals the true effect from the DGP independent of the selected TSVC model structure. 
 
\begin{table*}[!t]
\caption{Coverage proportions of 95\% CIs based on 5000 replications for a linear DGP (scenario 1). For each CI method the first row shows the coverage proportion averaged across all coefficients ($C_\text{av}$). Below coverage proportions are reported separately for each covariate ($C_j$). Coverage proportions for 90\% CIs are given in Table S1 in the Supplementary Material.}
\label{sc1CI95}
\begin{center}
\begin{footnotesize}
\begin{tabularx}{\textwidth}{l l l Y Y Y Y Y Y Y Y Y}
\toprule
CI method & Covariate & $n$ & \multicolumn{2}{c}{200} & \multicolumn{2}{c}{500} & \multicolumn{2}{c}{1000} \\
 & & $\sigma_\varepsilon$&1 & 2 & 1 & 2  &1& 2\\ \hline
Wald  & --- & & \textbf{.833} & \textbf{.833} & \textbf{.875} & \textbf{.875} & \textbf{.900} & \textbf{.900} \\
& $X_1$ & & .832 & .832 & .874 & .874 & .903 & .903 \\
 & $X_2$  & & .835 & .835 & .877 & .877 & .896 & .896 \\
Parametric percentile & ---  & & \textbf{.951} & \textbf{.951} & \textbf{.949} & \textbf{.949} & \textbf{.949} & \textbf{.949} \\
 & $X_1$  & & .950 & .950 & .949 & .949 & .951 & .951 \\
 & $X_2$  & & .952 & .952 & .949 & .949 & .947 & .947 \\
Bootstrap calibration & --- & & \textbf{.883} & \textbf{.883} & \textbf{.902} & \textbf{.902} & \textbf{.916} & \textbf{.916} \\
 & $X_1$ & & .880 & .880 & .900 & .900 & .919 & .919 \\
 & $X_2$ & & .885 & .885 & .903 & .903 & .914 & .914  \\
\bottomrule
\end{tabularx}	
\end{footnotesize}
\end{center}		
\end{table*}

From Table~\ref{sc1CI95} it is seen that the proposed parametric percentile CIs yielded coverage proportions very close to the nominal level across all settings (with varying $n$ and $\sigma_\varepsilon
$). The coverage proportions of the Wald type CIs were far too low but increased with higher sample size. This is likely due to the lower number of splits performed for larger samples (see Table~\ref{sc1Splits}) and the fact that the Wald type CIs are valid and yield the desired coverage if no splits are performed (i.e.\@ if the coefficients are simply non-varying). The bootstrap-calibrated CIs showed improved coverage proportions compared to the Wald type CIs but performed considerably worse than our proposed CIs (coverage proportions $<0.92$). Analogous results were observed for 90\% CIs (see Supplementary Material Table S1).

Of note, even with this simple underlying linear DGP without varying effects, neglecting the fact that constructing CIs for TSVCs is a selective inference problem (e.g. by applying a naive Wald type CI) may yield highly anti-conservative results with low coverage. 

\subsection{Varying effect DGP}

The data in the second scenario was generated by
\begin{equation}
\label{varyDGP}
y_{i} = 0.5\, I(x_{i2} \leq 0.5 \land x_{i3} = 1)\, x_{i1} -I(x_{i2} > 0.5)\, x_{i1} + \varepsilon_i\, ,\,  i = 1,\dots , n\, . 
\end{equation}
Here, a varying effect of $X_1$ that is determined by a tree structure with three terminal nodes was present. The covariates $X_2$ and $X_3$ did not have a linear effect but served only as effect modifiers for $X_1$. The proportions of variance explained by the covariate and the effect modifiers were $0.26$ ($\sigma_{\varepsilon}=1$) and $0.08$ ($\sigma_{\varepsilon} = 2$).

Table~\ref{sc2Splits} shows that the average number of splits performed by TSVC decreased with sample size and noise. It is also seen that the proportion of splits performed for the linear effect of $X_1$ increased with sample size, which suggests that the structure of the fitted models aligned more closely with the structure of the DGP if $n$ was large. Furthermore, the table shows that $X_2$ was more likely to be selected as an effect modifier than $X_3$, which reflects the tendency of the tree building towards the selection of continuous over binary splitting variables.

\begin{table*}[!t]
\caption{Average number of splits based on 5000 replications when fitting a TSVC model for a varying effect DGP (scenario 2). The first row shows the average total number of splits per replication in the TSVC models. Below the average number of splits per replication is reported separately for each combination of covariate and effect modifier. The column \textit{DGP} contains the true number of splits.}
\label{sc2Splits}
\begin{center}
\begin{footnotesize}
\begin{tabularx}{\textwidth}{l l ll Y Y Y Y Y Y Y Y Y}
\toprule
Covariate & Effect modifier & DGP & $n$ & \multicolumn{2}{c}{200} & \multicolumn{2}{c}{500} & \multicolumn{2}{c}{1000} \\
  & & & $\sigma_\varepsilon$&1 & 2 & 1 & 2  &1& 2\\ \hline
\textbf{ ---} & \textbf{---} & \textbf{2} & & \textbf{3.55} & \textbf{3.05} & \textbf{3.04} & \textbf{2.73} & \textbf{2.87} & \textbf{2.66} \\
$X_1$ & --- & \textbf{2} & & \textbf{2.47} & \textbf{1.43} & \textbf{2.42} & \textbf{2.00} & \textbf{2.47} & \textbf{2.22} \\
& $X_2$ & 1 & & 1.50 & 1.03 & 1.41 & 1.14 & 1.43 & 1.23 \\
& $X_3$ & 1 & & 0.97 & 0.40 & 1.02 & 0.86 & 1.03 &  1.00 \\
$X_2$ & \textbf{---} & \textbf{0} & & \textbf{0.31} & \textbf{0.42} & \textbf{0.18} & \textbf{0.20} & \textbf{0.12} & \textbf{0.13} \\
& $X_1$ & 0 & & 0.29 & 0.39 & 0.18 & 0.19 & 0.12 & 0.13 \\
& $X_3$ & 0 & & 0.02 & 0.03 & 0.01 & 0.01 & 0.00 & 0.01 \\
$X_3$ & \textbf{---} & \textbf{0} & & \textbf{0.77} & \textbf{1.20} & \textbf{0.43} & \textbf{0.53} & \textbf{0.28} & \textbf{0.30} \\
& $X_1$ & 0 & & 0.42 & 0.72 & 0.23 & 0.31 & 0.15 & 0.16 \\
& $X_2$ & 0 &
 & 0.35 & 0.48 & 0.21 & 0.22 & 0.13 & 0.14 \\
\bottomrule

\end{tabularx}	
\end{footnotesize}
\end{center}		
\end{table*}

\begin{figure*}[!t]
\centering
\includegraphics[width = 11.9cm]{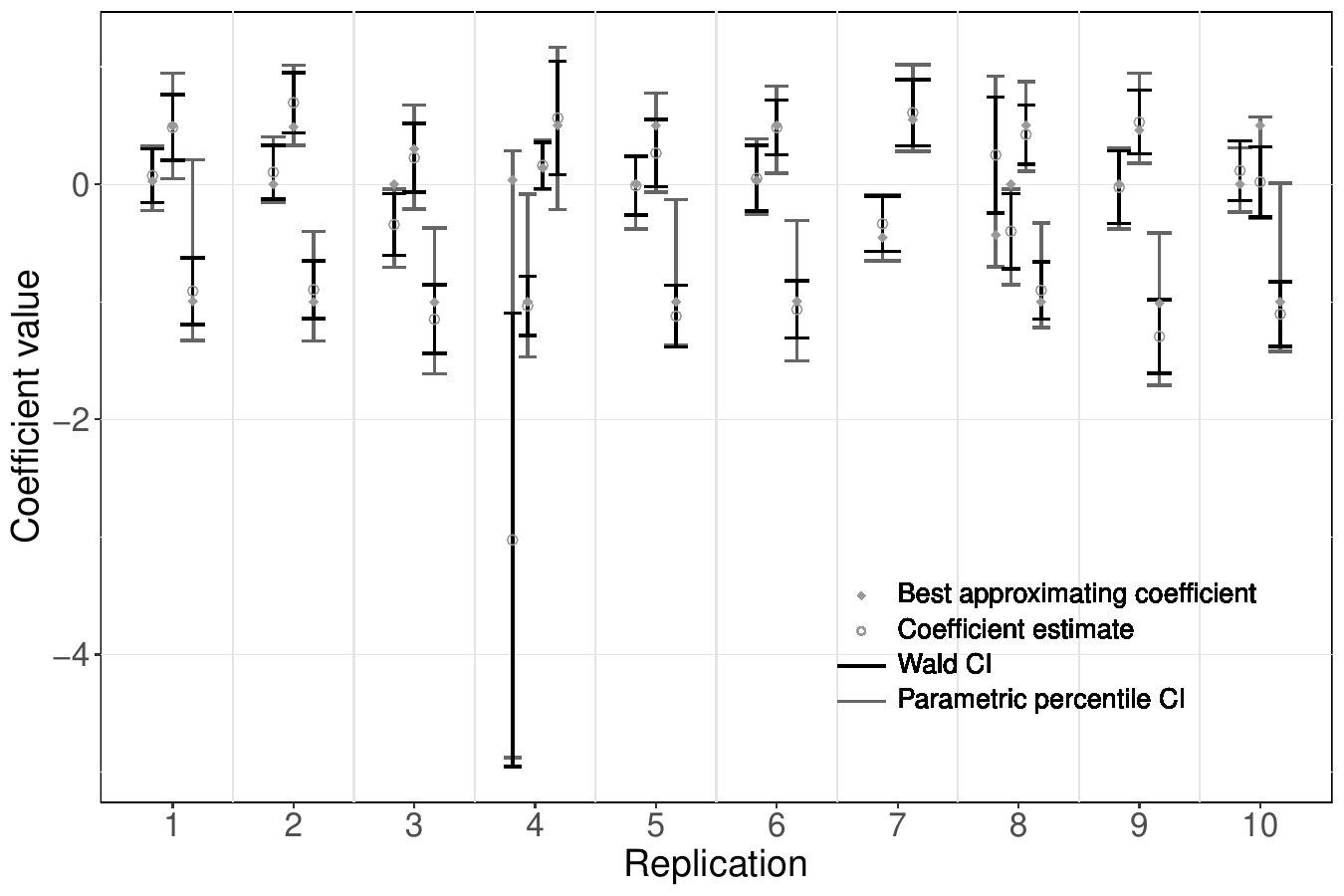}
\caption{Fitted tree-structured varying coefficients and 95\% CIs for a varying effect DGP (scenario 2). The figure shows the estimated effects $\hat{\beta}^{\mathcal{M}_r}_{jm}$of the original model fitted to the data $\mathcal{D}$, the best approximating linear coefficients $\beta^{\mathcal{M}_r}_{jm}$ as defined in Equation~\eqref{defBetaM}, corresponding 95\% Wald type and parametric percentile CIs for the varying linear coefficients of $X_1$ for 10 exemplary replications. The number of coefficients of $X_1$ in the true varying effect DGP is $3$. The underlying data were drawn from a linear DGP with $n = 200$ and $\sigma_{\varepsilon} = 1$.}
\label{CoverageSC2}
\end{figure*}

Figure~\ref{CoverageSC2} illustrates that the parametric bootstrap approach yielded wider CIs than the normal distribution-based Wald type approach (which resulted in much better coverage proportions, see Table~\ref{sc2CI95}). Note that, in this scenario, the best approximating coefficients differed between replications depending on the structure of the fitted TSVC model.  In most of the depicted replications the three regions with coefficients $\beta_{11} = 0$, $\beta_{12} = 0.5$ and $\beta_{13} = 1$ appear to be identified quite well. There were, however, also cases where too many (replications~4 and~8) or too few splits in the coefficient of $X_1$ were performed (replication 7).

Table~\ref{sc2CI95} shows that the coverage proportions of the proposed parametric percentile CIs were slightly conservative but approached the nominal level of $95\%$ for larger sample size and lower noise. Coverage proportions for the coefficients of $X_1$ exceeded the nominal level the most whereas the CIs for the coefficients of $X_3$ (eighth row in Table~\ref{sc2CI95}) were close to the nominal level across all settings. This may be due to the fact that $X_3$ was a binary variable and therefore allowed only one split point when selected as effect modifier. While the bootstrap calibration approach outperformed the normal distribution-based Wald type approach, both resulted in insufficiently low coverage across all settings ($<0.87$ and $<0.92$ on average). Coverage of the Wald type CIs increased with larger sample size, but no differences with regard to $n$ and $\sigma_\varepsilon$ were apparent for bootstrap calibration. The coverage proportions of the 90\% CIs exhibited an overall similar pattern (see Supplementary Material Table S2). 

\begin{table*}[!t]
\caption{Coverage proportions of 95\% CIs based on 5000 replications for a varying effect DGP (scenario 2). For each CI method the first row shows the coverage proportion averaged across all coefficients ($C_\text{av}$). Below coverage proportions are reported separately for each covariate ($C_j$). Coverage proportions for 90\% CIs are given in Table S2 in the Supplementary Material.}
\label{sc2CI95}
\begin{center}
\begin{footnotesize}
\begin{tabularx}{\textwidth}{l l l Y Y Y Y Y Y Y Y Y}
\toprule
CI method & Covariate & $n$ & \multicolumn{2}{c}{$200$} & \multicolumn{2}{c}{$500$} & \multicolumn{2}{c}{$1000$} \\
 & & $\sigma_\varepsilon$&1 & 2 & 1 & 2  &1& 2\\ \hline
Wald  & --- & & \textbf{.795} & \textbf{.795} & \textbf{.843} & \textbf{.852} & \textbf{.865} & \textbf{.867} \\
& $X_1$ & & .824 & .807 & .862 & .873 & .879 & .883 \\
 & $X_2$  & & .797 & .801 & .849 & .857 & .868 & .872 \\
 & $X_3$  & & .764 & .777 & .817 & .825 & .849 & .846  \\
Parametric percentile  & ---  & & \textbf{.968} & \textbf{.971} & \textbf{.966} & \textbf{.970} & \textbf{.963} & \textbf{.965} \\
 & $X_1$  & & .981 & .984 & .979 & .985 & .972 & .977 \\
 & $X_2$  & & .971 & .975 & .971 & .972 & .971 & .972 \\
 & $X_3$  & & .951 & .955 & .948 & .952 & .947 & .948 \\
Bootstrap calibration & --- & & \textbf{.901} & \textbf{.911} & \textbf{.901} & \textbf{.914} & \textbf{.911} & \textbf{.908} \\
 & $X_1$ & & .922 & .916 & .914 & .934 & .918 & .917 \\
 & $X_2$ & & .896 & .913 & .899 & .909 & .911 & .809 \\
 & $X_3$ & & .886 & .903 & .890 & .899 & .903 & .899 \\
\bottomrule
\end{tabularx}	
\end{footnotesize}
\end{center}		
\end{table*}

\subsection{Varying effect DGP with known effect modifiers}

In the third scenario, the varying effect DGP from Equation~\eqref{varyDGP} was applied again. However, compared to scenario~2 it is now assumed that the effect modifiers were known before model fitting, i.e. it was specified that only $X_2$ and $X_3$ are considered as potential effect modifiers in the TSVC fitting procedure. This type of scenario is common in applications, where prior knowledge is available or certain shapes of interactions between variables are scientifically not meaningful (see for example the application to real-world data in Section~\ref{subsec:app2}). Note that, while knowing the effect modifiers simplified the model selection problem, the TSVC algorithm still needed to detect which coefficients are modified by which effect modifier and the corresponding splitting rule.

\begin{table*}[!t]
\caption{Average number of splits based on 5000 replications when fitting a TSVC model for a varying effect DGP with known effect modifiers (scenario~3). The first row shows the average total number of splits per replication in the TSVC models. Below the average number of splits per replication is reported separately for each combination of covariate and effect modifier. The column \textit{DGP} contains the true number of splits.}
\label{sc3Splits}
\begin{center}
\begin{footnotesize}
\begin{tabularx}{\textwidth}{l l l l Y Y Y Y Y Y Y Y}
\toprule
Covariate & Effect modifier & DGP & $n$ & \multicolumn{2}{c}{200} & \multicolumn{2}{c}{500} & \multicolumn{2}{c}{1000} \\
 & & & $\sigma_\varepsilon$&1 & 2 & 1 & 2  &1& 2\\ \hline
\textbf{---} & \textbf{---} & \textbf{2} & & \textbf{2.95} & \textbf{2.39} & \textbf{2.64} & \textbf{2.45} & \textbf{2.60} & \textbf{2.37} \\
\textbf{$X_1$} & \textbf{---} & \textbf{2} & & \textbf{2.64} & \textbf{2.06} & \textbf{2.47} & \textbf{2.29} & \textbf{2.49} & \textbf{2.26} \\
 & $X_2$ & 1 & & 1.61 & 1.35 & 1.44 & 1.29 & 1.45 & 1.24 \\
 & $X_3$ & 1 & & 1.03 & 0.71 & 1.03 & 0.99 & 1.04 & 1.01 \\
$X_2$ & $X_3$ & 0 & & 0.01 & 0.01 & 0.01 & 0.01 & 0.00 & 0.00 \\
$X_3$ & $X_2$ & 0 & & 0.30 & 0.32 & 0.16 & 0.16 & 0.11 & 0.11 \\
\bottomrule
\end{tabularx}	
\end{footnotesize}
\end{center}		
\end{table*}

The average number of splits shown in Table~\ref{sc3Splits} was lower than in the second scenario and closer to the true number of 2 splits across all settings. As in the previous scenarios, fewer splits were performed if the sample size was large and the level of noise was high. In addition, it is seen that nearly none of the splits were performed in the coefficient of $X_2$, where the only available splitting option was~$X_3$.

Table~\ref{sc3CI95} shows that the proposed parametric percentile CIs yielded coverage proportions close to the nominal level for the coefficients of $X_1$ and $X_3$ (sixth and eighth row) and rather conservative coverage proportions for the coefficients of $X_2$ (seventh row) across all settings. The fact that $X_1$ was no longer considered as a potential effect modifier for $X_2$ and $X_3$ led to substantially improved coverage proportions for the coefficients of~$X_1$ compared to the results observed in the previous scenario (see Table~\ref{sc2CI95}). The Wald type and bootstrap-calibrated CIs again tended to be too short but yielded coverage proportions closer to nominal level of $95\%$ compared to scenario 2. The bootstrap-calibrated CIs achieved proportions around $0.93$ if the sample size was large and the level of noise was low. Similar results were observed for the 90\% CIs (see Supplementary Material Table S3).

\begin{table*}[!t]
\caption{Coverage proportions of 95\% CIs based on 5000 replications for a varying effect DGP with known effect modifiers (scenario 3). For each CI method the first row shows the coverage proportion averaged across all coefficients ($C_\text{av}$). Below coverage proportions are reported separately for each covariate ($C_j$). Coverage proportions for 90\% CIs are given in Table S3 in the Supplementary Material.}
\label{sc3CI95}
\begin{center}
\begin{footnotesize}
\begin{tabularx}{\textwidth}{l l l Y Y Y Y Y Y Y Y Y}
\toprule
CI method & Covariate & $n$ & \multicolumn{2}{c}{$200$} & \multicolumn{2}{c}{$500$} & \multicolumn{2}{c}{$1000$} \\
 & & $\sigma_\varepsilon$&1 & 2 & 1 & 2  &1& 2\\ \hline
Wald  & --- & & \textbf{.873} & \textbf{.870} & \textbf{.903} & \textbf{.903} & \textbf{.915} & \textbf{.915} \\
& $X_1$ & & .862 & .856 & .895 & .897 & .906 & .910 \\
 & $X_2$  & & .899 & .899 & .920 & .919 & .934 & .931 \\
 & $X_3$  & & .858 & .855 & .893 & .892 & .906 & .903  \\
Parametric percentile & ---  & & \textbf{.952} & \textbf{.957} & \textbf{.955} & \textbf{.957} & \textbf{.956} & \textbf{.956} \\
 & $X_1$  & & .948 & .961 & .951 & .955 & .949 & .951 \\
 & $X_2$  & & .970 & .970 & .971 & .972 & .972 & .972 \\
 & $X_3$  & & .939 & .940 & .944 & .943 & .946 & .946 \\
Bootstrap calibration & --- & & \textbf{.923} & \textbf{.928} & \textbf{.928} & \textbf{.930} & \textbf{.934} & \textbf{.931} \\
 & $X_1$ & & .925 & .925 & .928 & .928 & .929 & .928 \\
 & $X_2$ & & .929 & .935 & .927 & .936 & .946 & .944 \\
 & $X_3$ & & .914 & .924 & .918 & .927 & .925 & .922 \\
\bottomrule
\end{tabularx}	
\end{footnotesize}
\end{center}		
\end{table*}

\section{Applications}
\label{App}

To illustrate the proposed parametric bootstrap approach for constructing CIs for TSVC, two applications to real-world patient data were considered. The results are described in the following.

\subsection{Patients with COVID-19}

We considered data from a retrospective study in patients with PCR-confirmed COVID-19 that were admitted to the infectious disease department of the University Hospital Bonn between March 2020 and November 2021. A main objective of the study was to investigate the effect of treatment with the monoclonal antibody combination casirivimab/imdevimab (CVIV) on the need for oxygen support in the further course of the disease.
We analyzed data from $n=238$ patients hospitalized within five days after infection. For more details on the study, see \citet{Huebner2023}. The characteristics of the patients included in our analysis are: treatment with CVIV (0: no, 1: yes) and age in years. 

\citet{Huebner2023} analyzed the data using propensity score-weighted logistic regression. A need for oxygen support was shown to be significantly less frequent following treatment with CVIV (at error level $\alpha = 0.05$). Exploratory analyses indicated higher age as one of most relevant risk factors for requiring oxygen support in COVID-19 patients.

Our objective was to detect possible interactions between the treatment effect of CVIV and age. In order to do so, we fitted a logistic TSVC model with binary outcome `need for oxygen support' (yes/no) to the data, where the BIC was used to select the optimal number of splits and the maximal number of splits considered was $S=5$. Then we applied the proposed parametric bootstrap approach to obtain percentile CIs of the coefficients based on $B=1000$ bootstrap samples. For comparison, we also calculated asymptotic normal distribution-based Wald type CIs. 

The results in Table~\ref{covid} show that one split in the treatment effect with regard to age at split point 60 years was performed. According to the coefficient estimates, patients of age 60 years or younger benefited more from the CVIV treatment than patients older than 60 years. The Wald type CIs implied significant effects of age and CVIV treatment in both identified age groups at error level $\alpha = 0.05$. The proposed parametric percentile CIs were much wider and indicated a significant effect of age and of treatment with CVIV for the group of patients aged 60 years or younger but no significant treatment effect for patients older than 60 years.
\begin{table*}[!t]
\caption{Effect estimates, odds ratios, and 95\% CIs of the logistic TSVC model fitted to the COVID-19 data.}
\label{covid}
\begin{center}
\begin{footnotesize}
\begin{tabularx}{\textwidth}{l l Y Y Y Y}
\toprule
Covariate & Partition & $\beta$ & $\exp (\beta)$ & Wald type CI & Parametric percentile CI \\ \hline
Age & ---  & 0.008 & 1.008 & [1.006, 1.042] & [1.006, 1.051]  \\
CVIV & $\text{Age}\leq 60$ & -2.924 & 0.054 & [0.012, 0.164] & [0.000, 0.252] \\
& $\text{Age} >60$ & -0.986 & 0.373 & [0.174, 0.797] & [0.001, 8.629] \\
\bottomrule
\end{tabularx}	
\end{footnotesize}
\end{center}		
\end{table*}

\subsection{Patients with acute odontogenic infection}\label{subsec:app2}

In a second application, we analyzed data from a retrospective study investigating hospitalized patients with abscess of odontogenic origin conducted between 2012 and 2017 by the Department of Oral and Cranio-Maxillo and Facial Plastic Surgery at the University Hospital Bonn. Patients with an acute odontogenic infection suffer from pain, swelling, erythema and hyperthermia. If not treated at an early stage, such infections may spread into deep neck spaces and lead to perilous complications by menacing anatomical structures, such as major blood vessels, the upper airway and the mediastinum \citep{Biasotto2004}. The primary objective of the study was to identify risk factors that are associated with a prolonged length of stay (LOS) in the treatment of severe odontogenic infections. LOS was recorded in days ($t = 1,\dots , 18$).
Here data from 303 patients that underwent surgical treatment in terms of incision and drainage of the abscess were considered. Intravenous antibiotics were administered
during the operation and for the length of inpatient treatment. Further details on the study can be found in \citet{Heim2019}. Characteristics of the patients relevant for modeling were:
age in years, spreading of the infection focus into facial spaces (0: no, 1: yes), and the presence of diabetes mellitus type 2 (0: no, 1: yes). 

\citet{Puth2020} analyzed the data using a logistic discrete hazard model with tree-structured varying coefficients (see the Supplementary Material for more details on the logistic discrete hazard model). Specifically, they allowed for the coefficients of all covariates to be modified by $t$ (time since admission), which was considered as the only potential effect modifier. The number of splits performed was determined using a permutation test \citep{Berger2019}. Their findings indicated that the effect of diabetes is modified by $t$, where patients suffering from diabetes are much less likely to be discharged within the first four days since admission but after four days the effect of diabetes vanishes.

We analyzed the data analogously, except using BIC to obtain the optimal number of splits with a maximal number of splits of $S=5$, and were able to reproduce the fitted model structure. Then we applied Wald type CIs and the proposed parametric bootstrap approach  based on $B=1000$ bootstrap samples to obtain 95\% CIs of the coefficients.

\begin{table*}[!t]
\caption{Effect estimates, continuation ratios, and 95\% CIs of the logistic discrete hazard TSVC model fitted to the odontogenic infection data.}
\label{ondo}
\begin{center}
\begin{footnotesize}
\begin{tabularx}{\textwidth}{l l Y Y Y Y}
\toprule
Covariate & Partition & $\beta$ & $\exp (\beta)$ & Wald type CI & Parametric percentile CI \\ \hline
Age & ---  &-0.008 & 0.992 & [0.983, 0.999] & [0.962, 1.010]  \\
Spreading & --- & -0.939 & 0.391 & [0.255, 0.584] & [0.001, 0.670] \\
Diabetes & $t\leq 4$ & -2.438 & 0.087 & [0.004, 0.409] & [0.000, 0.725] \\
& $t >4$ & 0.002 & 1.000 & [0.578, 1.695] & [0.091, 4.910] \\
\bottomrule
\end{tabularx}	
\end{footnotesize}
\end{center}		
\end{table*}     

The results are shown in Table~\ref{ondo}.  The parametric percentile CIs were again much wider than the Wald type CIs and indicated a significant effect of spreading of the infection focus into facial spaces on the time to discharge but no significant effect of age (in contrast to the Wald type CI). Importantly, both CI methods indicated that diabetes significantly decreased the probability of being discharged within the first 4 days since admission, whereas no effect of diabetes after day 4 was shown, confirming the findings by \citet{Puth2020}.   

%This approach yields a sample of bootstrap estimates $\bar{\beta}^{(b)}_{jm}, b = 1,\dots, B$. Based on the percentile method a $(1-\alpha )$ confidence interval of $\tilde{\beta}_{jm}$ is constructed as
%\begin{equation}
%CI_{P} = \left [\bar{\beta}^{(\alpha/2)}_{jm}, \, \bar{\beta}^{(1-\alpha/2)}_{jm} \right]\, ,
%\end{equation}
%\begin{sloppypar}
%where $\bar{\beta}^{(q)}_{jm}$ denotes the $100q$th percentile of the bootstrap estimates $\bar{\beta}^{(b)}_{jm}, {b = 1,\dots, B}$.
%\end{sloppypar}

%, and that tree structure determining the coefficient of $x_j$ in the fitted model yields the subsets $\hat{N}_{jm}, m = 1,\dots, \hat{M}_{j}$ as leaf nodes. Then we define the true effect of leaf node $\hat{N}_{jm}$ as 
%\begin{equation}
%\bar{\beta}_{jm} = \frac{1}{\hat{n}_{jm}}\sum_{i\in \hat{N}_{jm}} \beta_{j}^{(i)}\, ,
%\end{equation}
%where $\hat{n}_{jm}$ denote the number of observations in $\hat{N}_{jm}$ and $\beta_{j}^{(i)}$ denotes the effect of $x_j$ assigned to observation $i$, i.e. 
%\begin{equation}
%\beta_j^{(i)}   = 
%\begin{cases}
%\beta_{j1}\text{, if }\bs{x}_{i[-j]} \in N_{j1} \\
%\quad \vdots \\
%\beta_{jM_j}\text{, if }\bs{x}_{i[-j]} \in N_{jM_j}
%\end{cases}
%\end{equation}
%With Equation~\eqref{trueEffect} a true effect for each node in the constructed tree structure can be determined, even in cases where the true underlying tree structure of an effect was not identified correctly.

\section{Summary and discussion}

TSVC models are flexible tools for generalized regression that allow the linear effects of the covariates to vary with the effect modifiers and apply a tree building procedure to inherently detect relevant effect modifiers. Constructing CIs for TSVCs is a selective inference problem as statistical inference is performed after model selection. In this vein, we proposed a parametric bootstrap approach tailored to the complex selection mechanism of TSVC. 

The applications to real-world data from COVID-19 patients and from patients suffering from acute odontogenic infection showed that the proposed CIs may differ strongly from naive Wald type CIs and lead to different conclusions when assessing statistical significance of the coefficients. Both, the effect of CVIV in the group of elderly patients and the effect of diabetes within the first four days of hospitalization are highly clinically meaningful. This highlights that accounting for the selective inference problem is essential when statistical inference on the parameters of a TSVC model is of interest. In the simulation study, our approach yielded coverage proportions close to the nominal level for the linear DGP whereas the simple Wald type CI and the bootstrap calibration approach by \citet{Loh2019} showed insufficient coverage. Low coverage proportions of bootstrap-calibrated CIs are also in line with findings in previous papers \citep{Neufeld2022}. The results of the simulation scenario $3$ (where the effect modifiers were prespecified before model fitting) also demonstrate that the performance of the CI methods depends on the complexity of the selection problem. In more complex scenarios, where the effect modifiers are not known beforehand (scenario $2$), the proposed approach showed rather conservative results for the coefficients of continuous covariates, whereas Wald type and bootstrap-calibrated CIs yielded coverage proportions that were far too low. 

The parametric bootstrap procedure offers an approximate solution for conditioning on the model selection event. As further refinement, weighted percentiles could be applied, where the bootstrap estimates from models with tree structures that are more similar to the originally fitted TSVC model are given more weight in the percentile calculation of the proposed algorithm (see step 4 in the algorithm described in Section~\ref{subsec:param}). 
%Such a weighting may, however, also lead to coverage rates below the nominal level in some cases (e.g. in scenarios with a linear DGP).

TSVC models can be fitted using the eponymous R add-on package \citep{Berger2018TSVC}. While the implementation generally allows that the effect of each covariate is modified by the other variables, the package also enables to flexibly incorporate prior knowledge about the model structure. For instance, if the effect modifiers are known beforehand (as in simulation scenario 3 and the application to the odontogenic infection data), this can be specified in the arguments of the TSVC modeling function. It is also possible to declare covariates having fixed non-varying linear effects, and covariates that serve as effect modifiers, only.   

\citet{Berger2019} proposed to apply permutation tests as an early stopping criterion for the tree building in TSVC models. In each iteration, a test is performed to assess whether a further split should be performed or not. That is, these tests allow for inference on the difference between the coefficients in two nodes but not on the coefficients themselves. Our approach fills this important gap and allows to quantify uncertainty of the parameter estimates and to assess statistical significance of the varying coefficients in each node. Of note, the proposed CI approach can be used in combination with any stopping criterion for the tree building in TSVC, including permutation tests and minimal node size tuning. Here, we selected the optimal number of splits based on the BIC. Alternatively, the Akaike information cirterion (AIC) or the cross-validated predictive log-likelihood can be applied.

The proposed CI approach may easily be extended to conduct inference on the probability estimates in the leaf nodes of a classification tree and parameters from other tree-based models (see for example \citealp{Zeileis2008}, \citealp{Schmid2016} and \citealp{Spuck2023}). It will be promising to further explore this in future research.

 \subsection*{Conflict of interests}
Declarations of interest:
None.

\subsection*{Acknowledgements}
Support by the German Research Foundation (DFG), grant BE 7543/1-1, is gratefully acknowledged.

\clearpage

\bibliographystyle{plainnat} % plainnat ist das natbib aequivalent zu plain 
\bibliography{bib_doi_new}

\clearpage

\begin{center}
{\large\bf SUPPLEMENTARY MATERIAL}
\end{center}

\subsection*{Bootstrap calibration}
Following Loh et al. (2019), to construct a bootstrap calibrated $100(1-\alpha )\%$ CI, we consider the equidistant sequence $\alpha_1 <\alpha_2 <... < \alpha_K = \alpha$, where we set $\alpha_1 = 0.0001$, $K = 200$ for $\alpha =0.1$ and $K = 100$ for $\alpha =0.05$. Then $B=1000$ bootstrap samples $\mathcal{D}_b = \{(y_{i}^{(b)}, \bs{x}_i^{(b)}), i = 1,\dots ,n\}$ are drawn by sampling with replacement from the original data $\mathcal{D}$, and on each bootstrap sample $\mathcal{D}_b$ a TSVC model $\mathcal{M}_b$ with coefficient estimates $\hat{\beta}_{jm}^{\mathcal{M}_b}$ is fitted. Around each of these estimates $K$ Wald type $100(1-\alpha_k)\%$ CIs $CI^{k}_W(\beta_{jm}^{\mathcal{M}_{b}})$ are constructed. Subsequently, a model $\mathcal{M}_{b0}$ with the fixed structure $\mathcal{M}_b$ is fitted to the original data, yielding the coefficient estimates $\hat{\beta}^{\mathcal{M}_{b0}}_{jm}$.  For $\alpha_k$ and covariate $X_j$ an average coverage rate with regard to the effect $\hat{\beta}^{\mathcal{M}_{b0}}_{jm}$ estimated on the original sample is given by
\begin{equation*}
\gamma_{jk} = \frac{1}{B}\sum_{b = 1}^{B} \frac{1}{M_j^b}\sum_{m = 1}^{M_j^b} I\left (\hat{\beta}_{jm}^{\mathcal{M}_{b0}} \in CI^{k}_W(\beta_{jm}^{\mathcal{M}_{b}})\right )\, ,
\end{equation*} 
where $M_{j}^{b}$ is the number of coefficients of $X_j$ in model $\mathcal{M}_b$. For each covariate $X_j$, we define $k_j$ as the smallest $k$ such that $\gamma_{jk}<1-\alpha$ and calculate the adjusted $\alpha$ levels as
\begin{equation*}
\alpha_{j}^{\text{adj}} = (1-f)\alpha_{k_j-1} + f\alpha_{k_j}\, ,
\end{equation*} 
where $f = (\gamma_{j(k_j-1)} -1 + \alpha)/(\gamma_{k_j -1} - \gamma_{k_j})$. We then construct a calibrated $100(1-\alpha)\%$ CI of $\beta_{jm}^{\mathcal{M}}$ based on the Wald type formula using the adjusted $\alpha$-levels
\begin{equation*}
CI_C(\beta_{jm}^{\mathcal{M}}) = \left[\hat{\beta}^{\mathcal{M}}_{jm} + z_{1-\alpha_j^{\text{adj}}/2}SE(\hat{\beta}^{\mathcal{M}}_{jm}), \hat{\beta}^{\mathcal{M}}_{jm} + z_{\alpha_j^{\text{adj}}/2}SE(\hat{\beta}^{\mathcal{M}}_{jm})\right] \, ,
\end{equation*} 
where $z_q$ denotes the $100q$-th percentile of the standard normal distribution.
\newpage

\subsection*{Additional results of the simulation study}

\begin{table*}[!h]
\caption{Coverage rates of 90\% CIs based on 5000 replications for a linear DGP (scenario 1). For each CI method the first row shows the coverage rate averaged across all coefficients ($C_\text{av}$). Below coverage proportions are reported separately for each covariate ($C_j$).}
\label{sc1CI90}
\begin{center}
\begin{footnotesize}
\begin{tabularx}{\textwidth}{l l l Y Y Y Y Y Y Y Y Y}
\toprule
CI method & Covariate & $n$ & \multicolumn{2}{c}{$200$} & \multicolumn{2}{c}{$500$} & \multicolumn{2}{c}{$1000$} \\
 & & $\sigma$&1 & 2 & 1 & 2  &1& 2\\ \hline
Wald  & --- & & \textbf{.771} & \textbf{.771} & \textbf{.823} & \textbf{.823} & \textbf{.847} & \textbf{.847} \\
& $X_1$ & & .768 & .768 & .822 & .822 & .849 & .849 \\
 & $X_2$  & & .774 & .774 & .823 & .823 & .845 & .845 \\
Parametric percentile & ---  & & \textbf{.901} & \textbf{.901} & \textbf{.898} & \textbf{.898} & \textbf{.900} & \textbf{.900} \\
 & $X_1$  & & .897 & .897 & .899 & .899 & .899 & .899 \\
 & $X_2$  & & .904 & .904 & .898 & .898 & .900 & .900 \\
Bootstrap calibration & --- & & \textbf{.814} & \textbf{.814} & \textbf{.848} & \textbf{.848} & \textbf{.863} & \textbf{.863} \\
 & $X_1$ & & .812 & .812 & .847 & .847 & .866 & .866 \\
 & $X_2$  & & .815 & .815 & .848 & .848 & .860 & .860 \\
\bottomrule
\end{tabularx}	
\end{footnotesize}
\end{center}	
\caption{Coverage rates of 90\% CIs based on 5000 replications for a varying effect DGP (scenario 2). For each CI method the first row shows the coverage rate averaged across all coefficients ($C_\text{av}$). Below coverage proportions are reported separately for each covariate ($C_j$).}
\label{sc2CI90}
\begin{center}
\begin{footnotesize}
\begin{tabularx}{\textwidth}{l l l Y Y Y Y Y Y Y Y Y}
\toprule
CI method & Covariate & $n$ & \multicolumn{2}{c}{$200$} & \multicolumn{2}{c}{$500$} & \multicolumn{2}{c}{$1000$} \\
 & & $\sigma$&1 & 2 & 1 & 2  &1& 2\\ \hline
Wald  & --- & & \textbf{.718} & \textbf{.714} & \textbf{.776} & \textbf{.787} & \textbf{.803} & \textbf{.806} \\
& $X_1$ & & .750 & .727 & .798 & .810 & .816 & .821 \\
 & $X_2$  & & .718 & .724 & .782 & .791 & .805 & .808 \\
 & $X_3$  & & .684 & .692 & .749 & .760 & .789 & .788  \\
Parametric percentile & ---  & & \textbf{.925} & \textbf{.935} & \textbf{.922} & \textbf{.932} & \textbf{.919} & \textbf{.924} \\
 & $X_1$  & & .952 & .961 & .945 & .960 & .935 & .941 \\
 & $X_2$  & & .928 & .941 & .930 & .936 & .926 & .932 \\
 & $X_3$  & & .894 & .905 & .890 & .900 & .896 & .898 \\
Bootstrap calibration & --- & & \textbf{.824} & \textbf{.836} & \textbf{.837} & \textbf{.858} & \textbf{.852} & \textbf{.850} \\
 & $X_1$ & & .848 & .825 & .846 & .882 & .860 & .857 \\
 & $X_2$ & & .821 & .791 & .840 & .854 & .851 & .852 \\
 & $X_3$ & & .803 & .773 & .822 & .837 & .846 & .841 \\
\bottomrule
\end{tabularx}	
\end{footnotesize}
\end{center}		
\end{table*}

\newpage
\begin{table*}[!h]
\caption{Coverage rates of 90\% CIs based on 5000 replications for a varying effect DGP with known effect modifiers (scenario 3). For each CI method the first row shows the coverage rate averaged across all coefficients ($C_\text{av}$). Below coverage proportions are reported separately for each covariate ($C_j$).}
\label{sc3CI90}
\begin{center}
\begin{footnotesize}
\begin{tabularx}{\textwidth}{l l l Y Y Y Y Y Y Y Y Y}
\toprule
CI method & Covariate & $n$ & \multicolumn{2}{c}{$200$} & \multicolumn{2}{c}{$500$} & \multicolumn{2}{c}{$1000$} \\
 & & $\sigma$&1 & 2 & 1 & 2  &1& 2\\ \hline
Wald  & --- & & \textbf{.804} & \textbf{.802} & \textbf{.844} & \textbf{.845} & \textbf{.857} & \textbf{.858} \\
 & $X_1$  & & .796 & .788 & .837 & .841 & .850 & .854 \\
 & $X_2$  & & .830 & .831 & .861 & .859 & .871 & .870 \\
 & $X_3$  & & .788 & .787 & .833 & .834 & .850 & .850  \\
Parametric percentile & ---  & & \textbf{.901} & \textbf{.911} & \textbf{.906} & \textbf{.907} & \textbf{.909} & \textbf{.909} \\
 & $X_1$  & & .898 & .924 & .902 & .906 & .902 & .902 \\
 & $X_2$  & & .926 & .931 & .930 & .929 & .930 & .931 \\
 & $X_3$  & & .877 & .880 & .887 & .886 & .896 & .895 \\
Bootstrap calibration & --- & & \textbf{.857} & \textbf{.866} & \textbf{.870} & \textbf{.877} & \textbf{.881} & \textbf{.877} \\
 & $X_1$ & & .854 & .860 & .868 & .876 & .877 & .873 \\
 & $X_2$ & & .862 & .876 & .880 & .884 & .890 & .887 \\
 & $X_3$ & & .849 & .864 & .862 & .880 & .876 & .871 \\
\bottomrule
\end{tabularx}	
\end{footnotesize}
\end{center}		
\end{table*}

\newpage

\begin{table*}[!h]
\caption{Average adjusted $\alpha$-levels for the bootstrap-calibrated CIs based on 5000 replications. The first row for each scenario and $\alpha_K$-level shows the adjusted $\alpha$-levels averaged across all variables. Below the average adjusted $\alpha$-levels are reported separately for each covariate.}
\label{calib}
\begin{center}
\begin{footnotesize}
\begin{tabularx}{\textwidth}{l l l l Y Y Y Y Y Y Y Y Y}
\toprule
Simulation scenario & $\alpha_{K}$ & Covariate & $n$ & \multicolumn{2}{c}{$200$} & \multicolumn{2}{c}{$500$} & \multicolumn{2}{c}{$1000$} \\
& & & $\sigma$ & 1 & 2 & 1 & 2  & 1& 2\\ \hline
1 & 0.05 & --- & & \textbf{.036} & \textbf{.036} & \textbf{.038} & \textbf{.038} & \textbf{.040} & \textbf{.040} \\
 & & $X_1$ & & .036 & .036 & .038 & .038 & .040 & .040 \\
 & & $X_2$ & & .036 & .036 & .038 & .038 & .040 & .040 \\
 & 0.10 & --- & & \textbf{.077} & \textbf{.077} & \textbf{.083} & \textbf{.083} & \textbf{.086} & \textbf{.086} \\
 & & $X_1$ & & .077 & .077 & .083 & .083 & .086 & .086 \\
 & & $X_2$ & & .077 & .077 & .083 & .083 & .086 & .086 \\ \hline
2 & 0.05 & --- & & \textbf{.017} & \textbf{.014} & \textbf{.028} & \textbf{.024} & \textbf{.031} & \textbf{.032} \\
 & & $X_1$ & & .016 & .013 & .028 & .025 & .031 & .340 \\
 & & $X_2$  & & .019 & .015 & .029 & .026 & .032 & .033 \\
 & & $X_3$  & & .016 & .013 & .026 & .022 & .029 & .029 \\
 & 0.10 & --- & & \textbf{.045} & \textbf{.038} & \textbf{.063} & \textbf{.058} & \textbf{.068} & \textbf{.071} \\
 & & $X_1$ & & .045 & .037 & .065 & .060 & .070 & .075 \\
 & & $X_2$ & & .048 & .040 & .066 & .060 & .070 & .073 \\
 & & $X_3$  & & .041 & .036 & .058 & .053 & 0.64 & .066 \\ \hline
3 & 0.05 & --- & & \textbf{.029} & \textbf{.024} & \textbf{.037} & \textbf{.033} & \textbf{0.038} & \textbf{.041} \\
 & & $X_1$  & & .025 & .020 & .035 & .032 & .036 & .040 \\
 & & $X_2$  & & .034 & .031 & .041 & .038 & .042 & .043 \\
 & & $X_3$  & & .027 & .021 & .036 & .030 & .037 & .040 \\
 & 0.10 & --- & & \textbf{.067} & \textbf{.058} & \textbf{.080} & \textbf{.074} & \textbf{.082} & \textbf{.085} \\
 & & $X_1$ & & .063 & .053 & .077 & .074 & .079 & .085 \\
 & & $X_2$ & & .074 & .069 & .085  & .081 & 0.87 & .088 \\
 & & $X_3$ & & .063 & .052 & .077 & .069 & .079 & .084 \\
\bottomrule
\end{tabularx}	
\end{footnotesize}
\end{center}	
\end{table*}

\clearpage

\subsection*{The logistic discrete hazard model}

If the event time was measured on a discrete scale, one may consider the \textit{discrete hazard function} given by
\begin{equation*}
\lambda ( t|\, \bs{X}) = P(T=t|\, T_i\geq t, \bs{X}),\, t = 1, \dots, k \, ,
\end{equation*}
where $T$ denotes the event time (i.e. the day of discharge from hospital in the apllication to the odontogenic infection data). That is, the conditional probability for the occurrence of an event at time $t$ given that the event has not occurred until $t$ is considered. For a fixed time $t$, the discrete hazard drives a binary variable that specifies whether an event occurs at time $t$  or not, given that $T \geq t$. Hence, strategies for modeling binary outcome variables can be adapted to discrete hazard modeling.

Specifically, the \textit{logistic discrete hazard model} is widely applied. It is given by
 \begin{equation*}
\lambda(t\vert\ \bs{X})=h(\eta (t, \bs{X})),\quad t = 1,...,k-1\ ,
\label{hazard1}
\end{equation*}
where $h(\cdot)$ is the logistic link function and $\eta(\cdot)$ is the predictor function depending on the covriates and time. A commonly assumed semi-parametric form of the predictor function is given by 
\begin{equation*}
\eta (t, \bs{X}) = \gamma_{0}(t) + \bs{X}^\top \bs{\beta}\ ,
\label{parametric}
\end{equation*}
where $\gamma_0 (t)$ describes the hazard over time (for any given values of $\bs{X}$, called \textit{baseline hazard}), and where the effects $\bs{\beta}$ of the explanatory variables on the hazard are assumed to be linear and independent of time (Tutz \& Schmid, 2016).  This model is also called the \textit{proportional continuation ratio model}. The continuation ratio at time $t$ is given by
\begin{equation*}
\Psi (t\vert\ \bs{X}) = \frac{P(T = t\vert\ \bs{X})}{P(T > t\vert\ \bs{X})} = \exp \left ( \gamma_{0} (t) + \bs{X}^\top \beta \right  )
\end{equation*}
and denotes the ratio comparing the probability of an event at time $t$ to the probability of an event after $t$. The expression exp($\beta_j$) corresponds to
\begin{equation*}
\exp ( \beta_j) = \frac{\Psi(t|\, X_j +1)}{\Psi(t|\, X_j)}\, ,
\end{equation*}
i.e. the ratio of continuation ratios. Hence, proportionality is given in the sense that the comparison of two individuals with regard to their continuation ratios is independent of time.

\end{document}